\documentclass[lettersize,journal]{IEEEtran}
\usepackage{amsmath,amsfonts}
\usepackage{algorithmic}
\usepackage{algorithm}
\usepackage{array}
\usepackage[caption=false,font=normalsize,labelfont=sf,textfont=sf]{subfig}
\usepackage{textcomp}
\usepackage{stfloats}
\usepackage{url}
\usepackage{verbatim}
\usepackage{graphicx}
\usepackage{cite}
\usepackage{threeparttable} 
\usepackage{multirow}
\usepackage[dvipsnames]{xcolor}
\usepackage{tabularx}

\makeatletter
\def\ps@IEEEtitlepagestyle{%
  \def\@oddfoot{}%
  \def\@oddhead{\mycopyrightnotice}%
  \def\@evenhead{\@IEEEheaderstyle\thepage\hfil\leftmark\hbox{}}\relax
  \def\@evenfoot{}%
}
\def\mycopyrightnotice{%
  \begin{minipage}{\textwidth}
  \centering \scriptsize
  This work has been submitted to the IEEE for possible publication. Copyright may be transferred without notice, after which this version may no longer be accessible.
  \end{minipage}
}
\makeatother

\begin{document}

\title{Evaluation of Mobile Network Slicing \\ in a Logistics Distribution Center}

\author{David Segura, Emil J. Khatib,~\IEEEmembership{Member,~IEEE,} Raquel Barco
\thanks{The  authors  are  with  Telecommunication  Research  Institute  (TELMA), Universidad  de  Málaga,  E.T.S.  Ingeniería  de  Telecomunicación,  Bulevar Louis  Pasteur  35,  29010,  Málaga  (Spain)  (e-mail: dsr@ic.uma.es, emil@uma.es, rbm@ic.uma.es). (Corresponding author: Emil J. Khatib.)}
}



\maketitle

\begin{abstract}
Logistics is a key economic sector where any optimization that reduces costs or improves service has a great impact on society at large. In this paper, the role of two 5G Network Slicing (NS) strategies in Smart Logistics is studied: the use of a static slice with a balance division of network resources and the use of a dynamic slice. To validate the potential gains of these strategies, a Distribution Center with 5G connectivity is simulated, recreating the activity that takes place in a real Smart Logistics scenario. Results show that a dynamic slice makes a more efficient usage of the network resources, improving the quality of service for the different traffic profiles, even when there is a traffic peak. This improvement ranges from 6.48\% to 95.65\%, depending on the specific traffic profile and the evaluated metric.
\end{abstract}


\section{Introduction}

\IEEEPARstart{I}n the last years, the emergence of networks and mobile communications has led to the development of new solutions that have revolutionized logistics. Wireless networks are one of the key enablers of Smart Logistics \cite{ding2021smart}, which allow the supply of products in small or individual batches with Just-In-Time delivery, reverse logistics, continuous feedback to clients, etc. Distribution centers \cite{distributioncenters} are a key element in the Smart Logistics supply chain, replacing traditional warehouses with lean nodes that act more as post offices where products spend few hours before being shipped in the next transport mean.

To support the Smart Logistics supply chain, there are numerous Industry 4.0 applications. For instance, Automated Guided Vehicles (AGVs \cite{agvs}) are used within distribution centers to move packages between different points; and Smart Tags \cite{smarttags} are used to track parcels at all times within a distribution center or even during transportation between different centers. The different applications running in Smart Logistics will have different requirements depending on the criticality of the messages, their size and the number of devices transmitting a message simultaneously. In the fifth-generation (5G) of cellular networks, three main traffic profiles are defined \cite{trafficprofiles}:
\begin{itemize}
    \item Enhanced Mobile Broadband (eMBB): messages that require a high bandwidth. Typically associated with multimedia applications, such as Augmented and Virtual Reality (AR/VR \cite{arvr}). 
    \item Ultra Reliable Low Latency Communications (URLLC): messages that require a very high reliability and very low latency. Normally, mission critical messages belong to this category, such as AGV navigation systems.
    \item Massive Machine Type Communications (mMTC): short, infrequent messages with low requirements for reliability and latency, but with a massive density of devices and a need for low power consumption. Applications such as Smart Tags belong to this category.
\end{itemize}

These traffic profiles are all present in Smart Logistics \cite{khatib2021optimization}. To allow such profiles with conflicting requirements to coexist on a single wireless network, 5G introduces Network Slicing (NS \cite{trafficprofiles}). With NS, the resources (physical machines, software, radio spectrum, etc.) are divided dynamically into independent sets with optimized configurations. A slice for each class of service can then be defined such that its requirements are met without negatively affecting other service classes.

Ranging from high-level studies of wireless applications in Smart Logistics \cite{ding2021smart, song2020applications}, to specific use-cases \cite{9552545,LI2021103565,9482459}, the topic of the application of 5G technologies on logistics is gaining an increasing interest from the research community. Several studies \cite{khatib2021optimization,ZHAN2022102052,9328620, DL-Anomaly-SmartLogistic, cheng2022-5GManufacturing-Review, URLLC-EMBB-Industrial-Survey} are centered in optimizing 5G networks specifically for Smart Logistics, analyzing the specific particularities of the applications and the different environments where the processes take place.

The use of NS \cite{NS-Overview-5G} has been widely agreed as a promising technique to accommodate diverse services that occur in industrial scenarios, such as distribution centers. In \cite{NS-Survey-SmartFactory}, the architecture and requirements of NS for smart factory is presented, along with the different challenges and implementation aspects. In \cite{NS-Evaluation-L5GO-MNO}, the authors perform an evaluation of the bandwidth utilization and the number of connected users, using different NS strategies. A proactive NS for Smart Logistics is proposed in \cite{khatib2021optimization} to adapt the radio resources into the different slices. This method is based on a Big Data prediction module to predict the traffic within a base station and then divide the resources proportionally among the expected traffic profiles.

This paper provides a practical study of network optimization in Logistics scenarios. To the best of the authors knowledge, there are no practical studies covering the optimization of 5G technology in logistics scenarios with simulations. The use of 5G technology in logistics has been previously proposed at a high level, but without providing network performance results under this particular scenario for the different traffic profiles (eMBB, URLLC and mMTC). Therefore, to cover this need, this paper provides a novel simulator which has been developed with a realistic representation of a distribution center, where several different logistics activities are done. The communications of these activities have been modeled and used to estimate the performance of the different traffic profiles. More specify, this paper tries to fill the gap of the network optimization methodology presented in \cite{khatib2021optimization} by providing an evaluation of the expected gains on the quality of service when using a dynamic slice compared to a static slice.

The remainder of this paper is organized as follows. In Section~\ref{Section-Background} a description of Smart Logistics along with its main scenario and applications is given. In Section~\ref{Section-WirelessConnectivity} a brief description of wireless connectivity in Industry 4.0 is given. In Section~\ref{Section-5G-Technologies}, the different 5G technologies that allow optimizing the network are explained. In Section~\ref{Section-Scenario}, the floorplan of the simulated distribution center is described. The simulator along with network parameters are described in Section~\ref{Section-Simulator}. The results are shown in Section~\ref{Section-Results} and discussed in Section~\ref{Section-Discussion}; and finally, the conclusions and future work are summarized in Section~\ref{Section-Conclusions}.

\section{Background}
\label{Section-Background}

Smart Logistics \cite{ding2021smart} emerges from the adoption of lean principles and the application of advanced Information Technology (IT) systems, with the objective of responding to new demands from the public. Compared to traditional logistics, Smart Logistics optimizes the processes for small product batches, where economies of scale cannot be applied easily; reduces the delivery times and allows reverse logistics. These principles apply to the full logistics chain, from the manufacturer to the consumer. To achieve this agility, the whole structure of distribution centers, transport vehicles and control and monitoring systems is changed with respect to traditional logistics. The need to support very small batches in an agile manner requires that in Smart Logistics, each parcel, delivery vehicle, storage facility resource, etc., is tracked in real time with a much higher precision, to allow for better planning. Such a sophisticated monitoring and control system can only be achieved with wireless connectivity, especially with cellular networks such as 5G, that allow connectivity within the distribution center and along the whole logistics chain.

\begin{figure}[ht]
\centering
\includegraphics[width=\columnwidth]{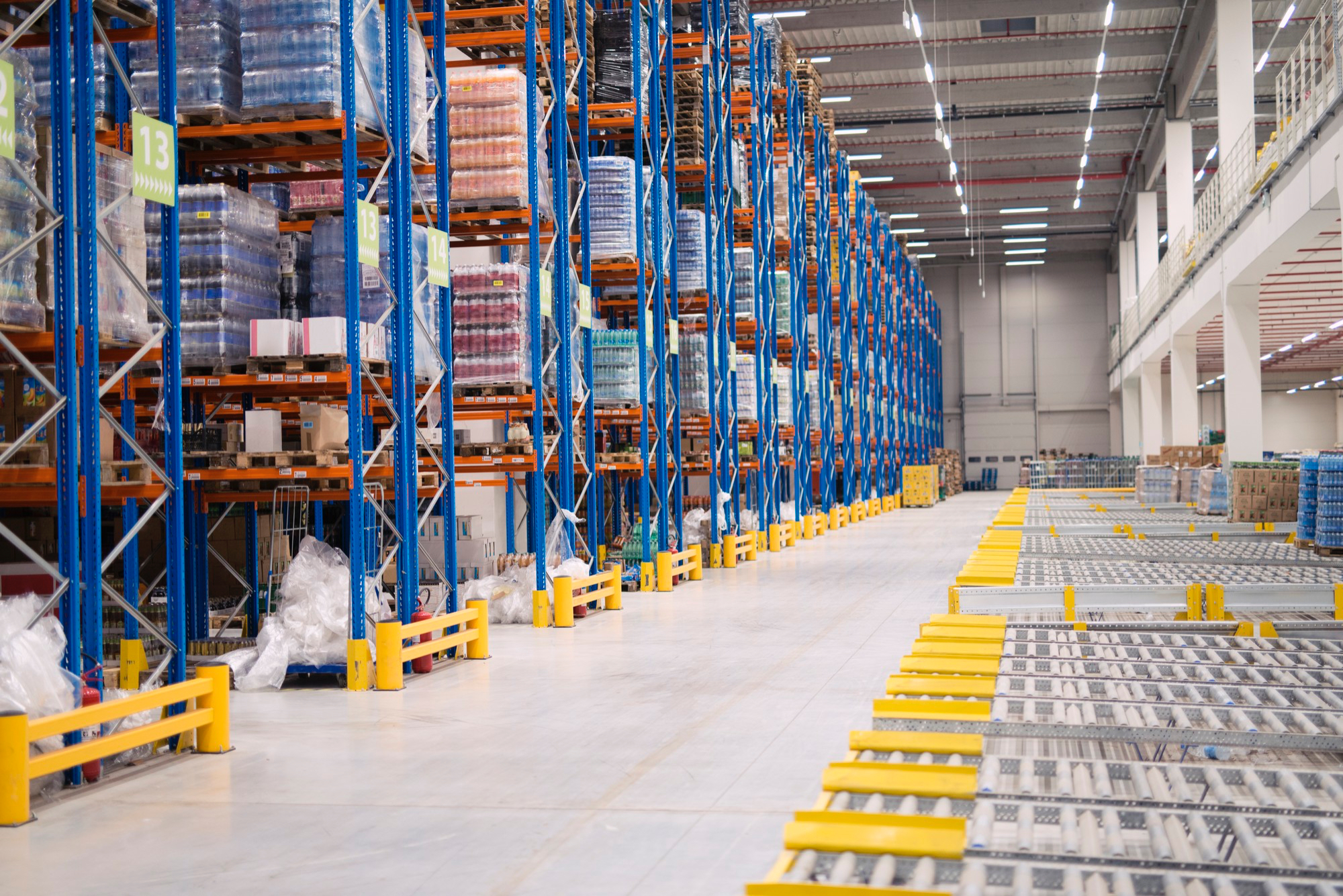}
\caption{Example of a distribution center.}
\label{5G.dc}
\end{figure}

The main scenario in Smart Logistics are the distribution centers \cite{distributioncenters}, which are large buildings with two main differentiated areas: the receiving and shipping docks (which may be the same or independent), and the storage area. The receiving and shipping docks are areas adjacent to doors adapted for loading and unloading trucks, where the equipment and personnel work to unload incoming and load outgoing goods. In Fig.~\ref{5G.dc} this area can be seen in the right side, with the door for loading/unloading trucks in the background. 

Advanced systems such as palletizing machines, AGVs \cite{agvs} and AR assistance for workers \cite{arvr} will operate in these areas. The storage area (left side in Fig.~\ref{5G.dc}) consists mainly of racks where products are stocked for short periods of time between arriving and being redirected to the next means of transport. In this area, products will be localized with systems such as Smart Tags \cite{smarttags} and stored/retrieved with AGVs.

The different applications that will run within a distribution center will have different requirements. For instance, AGV communications can be considered an URLLC profile, since large delays may cause malfunctions such as collisions between different vehicles or even with human workers. For this reason, reliability must also be very high and can be improved with multi-connectivity and packet duplication approach \cite{segura2022dynamic}. AR/VR \cite{remoteassistance} will require the uplink transmission of live video feeds and downlink transmission of complex 3D objects, so they will need a very high bandwidth (up to 50 Mbps) and ideally an end-to-end latency (including processing of 3D objects) below 20 ms to avoid dizziness \cite{arvr}. This matches the eMBB traffic profile. Smart Tags, on the other hand, have the main priority of conserving energy and having a very high coverage to allow connectivity within very cluttered environments (such as racks or pallets of parcels), with low bandwidth, latency and reliability requirements.

In a distribution center, there will be some particularities that differentiate the wireless network from other scenarios. First, the combination of services depends greatly on the activity taking place at a certain time in the center. For instance, when unloading an incoming container, the receiving dock will be populated by workers (possibly wearing and making active use of AR/VR glasses) and AGVs with URLLC control messages assisting them. After that period, products will be moved to specific places in the storage, again with a combination of workers and AVGs. A similar activity will be present when loading containers. Between loading and unloading (or in areas that are far from the docks in large distribution centers), only Smart Tags will be active sending periodic readings, with the occasional worker or AGV passing by. Smart Tags will be active at all times, creating a massive background traffic with low priority, but with tight energy and coverage requirements. Second, the high clutter caused by racks, machines and parcels will make a distribution center a very harsh environment for propagation, similar to the conditions found in a factory \cite{segura20215g}.

To provide connectivity for these applications, several wireless solutions are available, as it will be explained in the next section. Nevertheless, only 5G supports all traffic profiles that occur in a distribution center, using techniques such as NS.

\section{Wireless connectivity in Industry 4.0}
\label{Section-WirelessConnectivity}

Traditionally, wired connections have been used in industrial networks to connect different elements such as Programmable Logic Controllers (PLC) \cite{PLC}, that is, the computers that control the machines, with each other or with the Manufacturing Execution System (MES) \cite{MES}. Process monitoring as well as alarm monitoring are usually contained in the MES, constituting an interface between the PLCs and the Enterprise Resource Planning (ERP) \cite{ERP}, which allows a global coordination at executive level.

With the arrival of the Industry 4.0 paradigm \cite{lasi2014industry}, it is intended to obtain more flexibility and the support of new use-cases. To provide this, wireless connectivity has gain an increasing usage in factories, which can also reduce the installation costs.

Wireless technologies are divided into two types of networks that enable different applications: Local Area Networks (LAN) and Wide Area Networks (WAN). LANs have a coverage range of up to 100 meters and can cover an area of a distribution center or many rooms. On the other hand, WANs provide a higher coverage range, from distances of a few kilometers up to whole countries.

Regarding LANs, the IEEE 802.11 family (WiFi) \cite{IEEE802.11-1, IEEE802.11-2} is the most common technology used in some industrial deployments, due to its low cost and wide availability of components. Also, customized solutions for factories and based on IEEE 802.15.4 have been used, such as WirelessHART \cite{WirelessHART}, WIA-PA \cite{WIA-PA}, ZigBee \cite{Zigbee}, ISA100.11a \cite{ISA100.11a} and IETF 6LoWPAN \cite{rfc4944-6LoWPAN}. On the other hand, regarding WANs, technologies such as SigFox \cite{SigFox} and LoRaWAN \cite{LoRaWAN} have been used in manufacturing plants for power limited devices. 

Moreover, cellular networks (GSM/GPRS, Long Term Evolution, LTE) have also been applied for specific applications, such as sensors \cite{GPRS-Sensor} and robotics \cite{LTE-Robotics}. These cellular networks were not design for Industrial Internet of Things (IIoT). Therefore, technologies such as NB-IoT, Cat-M1 and EC-GSM \cite{MMTC-Technologies} were designed, focusing on energy saving and high coverage range with a low data rate.

The main problem of the non-cellular technologies mentioned above is that they are insufficient for logistics needs, since they are either not wide area networks, which limits the interoperability, or they do not offer the necessary data transmission capacity. Likewise, the actual technologies based on cellular networks such as NB-IoT, Cat-M1 and EC-GSM only fulfills the needs of coverage. Moreover, although LTE was designed as a broadband access technology, it cannot fulfill the requirements of extreme industrial applications, such as AR/VR. Apart from that, the latency and high reliability requirements for the new use-cases in factories (i.e., AGVs navigation or closed loop control) cannot be achieved with LTE. In other words, none of these technologies is capable of providing a service that covers the three main traffic profiles that occur in a distribution center (eMBB, URLLC and mMTC).

Only 5G supports all the traffic profiles, using techniques such as NS \cite{khatib2021optimization}. Since 5G is a WAN, it support communications over the whole logistics chain, not limited to the distribution center. With Non-Public Networks \cite{5G-Non-public-network}, the logistics operators can have custom connectivity with a private network supported over a public operator hardware, with the corresponding reduction in ownership and expertise costs.

\section{5G technologies}
\label{Section-5G-Technologies}

\subsection{Numerologies in 5G}
\label{Section-Numerologies}

5G New Radio (NR) introduced in Release 15 a new frame structure to provide flexibility and the adoption of new use-cases such as critical communications. The frame structure~in 5G NR can adopt different numerologies. A numerology ($\mu$) is defined by a cyclic prefix (normal or extended) and a SubCarrier Spacing (SCS). 

Four new numerologies have been defined in 5G NR, which ranges from 0 to 4, where $\mu = 0$ corresponds to LTE configuration. A numerology defines the SCS as $15\cdot2^{\mu}$~kHz and the slot duration~as~$1/2^{\mu}$~ms. As numerology increases, shorter slots are used, but they are wider in frequency, so a higher SCS is necessary. The standard states that not all numerologies are suitable for a determined frequency range (FR) and its use is divided into synchronization and data channels \cite{3GPP-21915}. 

\begin{figure}[ht]
\centering
\includegraphics[width=\columnwidth]{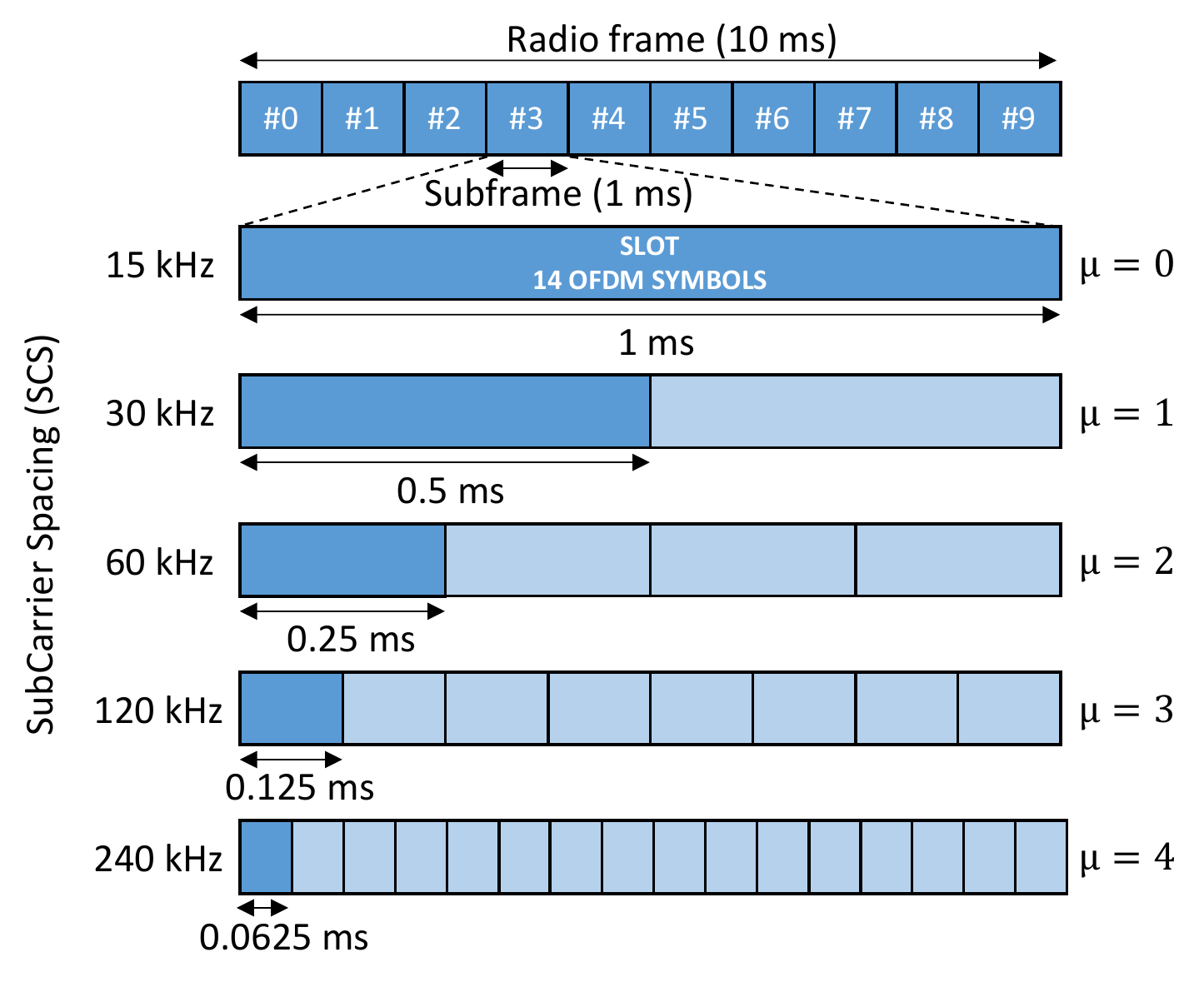}
\caption{5G numerologies scheme in the time domain.}
\label{5G.NumerologyScheme}
\end{figure}

For synchronization channels, $\mu = \{0, 1\}$ is used in FR1 (sub-6~GHz bands) and $\mu=\{3, 4\}$ in FR2 (millimeter wave bands). On the other hand, for data channels, $\mu=\{0, 1, 2\}$ is supported in FR1 and $\mu=\{2, 3\}$ in FR2. The number of subcarriers in 5G NR is 12 for all numerologies. Same as LTE, in 5G NR the frame duration is fixed at 10~ms and subframe duration at 1~ms. Depending on the selected configuration, the number of slots per subframe is defined as $2^\mu$. Finally, one slot is composed by 14 Orthogonal Frequency Division Multiplexing (OFDM) symbols, where the symbol duration is defined as $1/(14\cdot2^\mu)$~ms. Fig.~\ref{5G.NumerologyScheme} shows a summary of the characteristics for each numerology in the time domain.

The numerology is one of the main accepted solutions to reduce the latency, since the slots are shorter as $\mu$ increases. 
The scheduler normally works at slot level, so a decrease in the slot duration makes the resource allocation faster. However, this comes at the cost of network efficiency, since lower numerologies are better for capacity (eMBB traffic), that is, throughput. Also, bandwidth and packet size are two factors that affect when selecting the numerology. If the numerology selected is very high and the packet size is larger, this can lead to an unexpected increase in latency \cite{segura20215g}, especially in poor radio conditions.

\subsection{Network slicing}
\label{Section-NS}

The different types of traffic profile (URLLC, eMBB and mMTC) have different requirements that can be achieved by optimizing network parameters; but these optimizations may cause conflicts. This is the case of numerology, as explained in Section \ref{Section-Numerologies}. To solve this problem, NS \cite{trafficprofiles,khatib2021optimization} has been proposed to assign one slice per type of traffic; each slice being optimized independently without affecting others.

\begin{figure}[ht]
\centering
\includegraphics[width=\columnwidth]{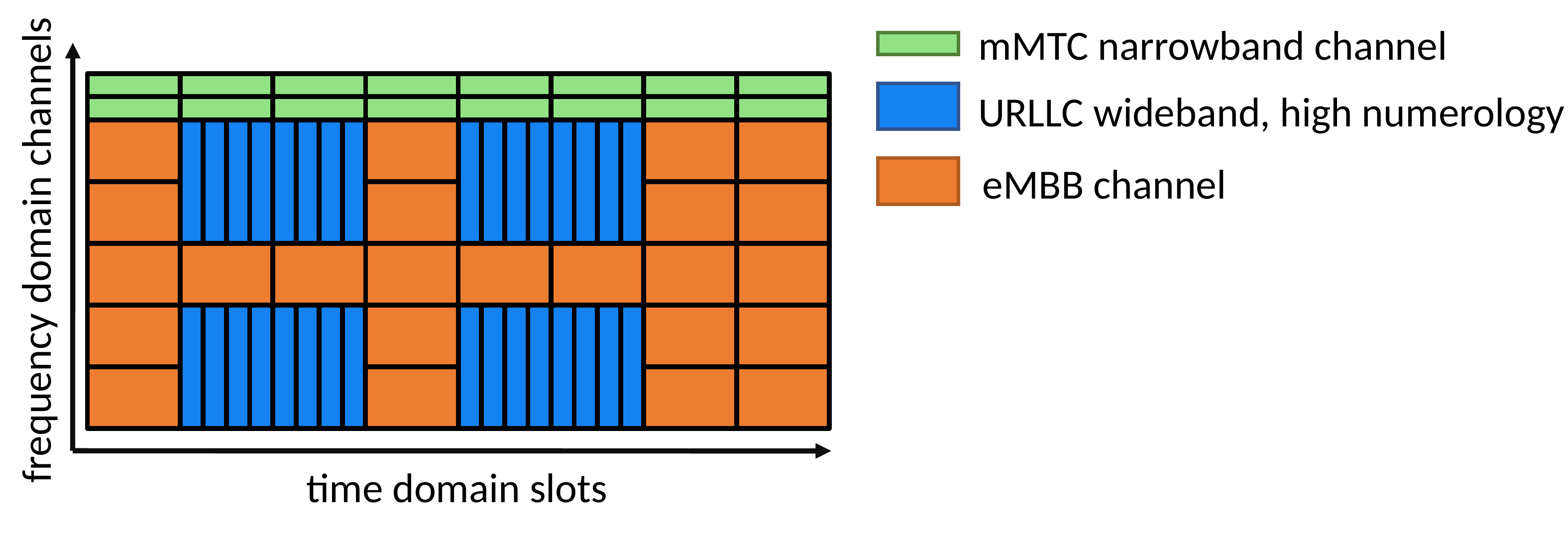}
\caption{NS for different service types.}
\label{N.S.ns}
\end{figure}

In NS, the carrier is divided into distinct frequency and time blocks, as shown in Fig.~\ref{N.S.ns}. Resources are then assigned to each type of service attending to their needs. For instance, the resources assigned to the URLLC service will use a high numerology, with the possibility of using redundant channels in time and frequency, while mMTC will use narrowband channels with low numerology to better adapt to poor radio conditions. The eMBB slice will use wideband channels and opportunistically reuse frequencies that are not used by URLLC.

While NS can be used in this manner to provide optimal channels to all services, resources may be wasted if they are not assigned proportionally. In these cases, NS allows to dynamically reshape the resource assignation. This can even be done proactively, using external data sources to predict when a specific profile will need more resources. In distribution centers \cite{khatib2021optimization}, data sources such as the truck schedule (combined with traffic information), the registry of online transactions or the current inventory can be used to estimate the composition of the traffic and therefore assign resources to the different profiles.

\subsection{Random access procedure}
\label{Section-RA}

The random access (RA) procedure is used for the User Equipments (UEs) to start communicating with the base station in cellular communications. It was first introduced in Release~8 and updated for 5G NR in Release 15. To handle this communication, two different RA procedures are defined in the standard \cite{3GPP-38300}:

\begin{itemize}
    \item Contention-based: where the UEs selects a randomly preamble from a pool of preambles to request a network access. This procedure is susceptible to collisions, therefore, it is used for delay-tolerant access (e.g., for the initial Radio Resource Control, RRC, connection establishment). 

    \item Contention-free: where the base station allocates dedicated resources (i.e., dedicated preamble) to the UEs, avoiding a preamble conflict. The preamble is allocated via RRC signaling or physical layer, and this procedure is used for delay-constrained access that requires a high probability of success (e.g., starting communication with the target base station in handover).
\end{itemize}

\begin{figure}[ht]
\centering
\includegraphics[width=0.75\columnwidth]{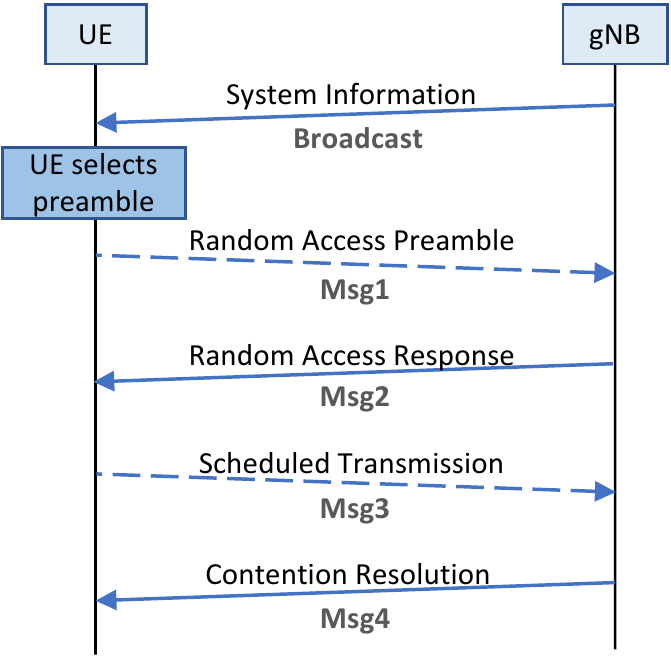}
\caption{Contention-based RA procedure.}
\label{CB-RA-procedure}
\end{figure}

In the contention-based RA procedure, there are four messages involved between the UE and the next generation NodeB (gNB), as depicted in Fig.~\ref{CB-RA-procedure}:

\begin{itemize}
    \item Msg1: the UE sends through the Physical Random Access Channel (PRACH) a preamble randomly selected among a list broadcast periodically by the gNB in the System Information Block (SIB) \cite{3GPP-38321, 3GPP-38331}. There are 64 preambles available for the RA, but not all of them are available for contention-based (some of them are reserved for contention-free access). A collision will occur in case that multiple UEs transmit the same preamble in the same RA slot. Once the UE sends the preamble, it waits a time window to receive a response from the gNB (Msg2). The duration of this window is broadcast by the gNB and has a maximum value of 10 ms \cite{3GPP-38331}. In case that the timer expires, the UE performs a new access attempt if the number of attempts is less than \textit{preambleTransMax}, which defines the maximum allowed value \cite{3GPP-38331}.

    \item Msg2: once the Msg1 is received at the gNB, it replies with a Random Access Response (RAR) message over the Physical Downlink Shared Channel (PDSCH) with a RA-RNTI and a temporary C-RNTI, providing an uplink resource grant and time alignment to be used to transmit the next message (Msg3) \cite{3GPP-38321}. The RA-RNTI identifies the preamble sent in Msg1, so the UE that transmitted that preamble is informed that it has been heard; and the C-RNTI is used by the UE to identify itself in the next steps. If the same preamble was selected by two or more UEs, the collides UEs will wait a random backoff time (according to the Backoff Indicator parameter, BI, attached to the RAR) before retrying a new access attempt (Msg1).

    \item Scheduled Transmission (Msg3): the UE starts sending it request over the Physical Uplink Shared Channel (PUSCH) along with the temporary C-RNTI \cite{3GPP-38300}, using the grant received on Msg2. The signaling message and the information associated will vary depending on the particular request: initial RRC connection setup, reestablishment of the RRC connection, etc.

    \item Contention Resolution (Msg4): upon reception of the connection request, the gNB replies with a contention resolution message. If an UE does not receive this message, it declares a contention resolution failure and the UE will perform a new access attempt, as previously explained in Msg1. If the counter reaches the maximum value (\textit{preambleTransMax}), the UE will indicate a random access failure to the upper layers.
    
\end{itemize}

\begin{figure}[ht]
\centering
\includegraphics[width=\columnwidth]{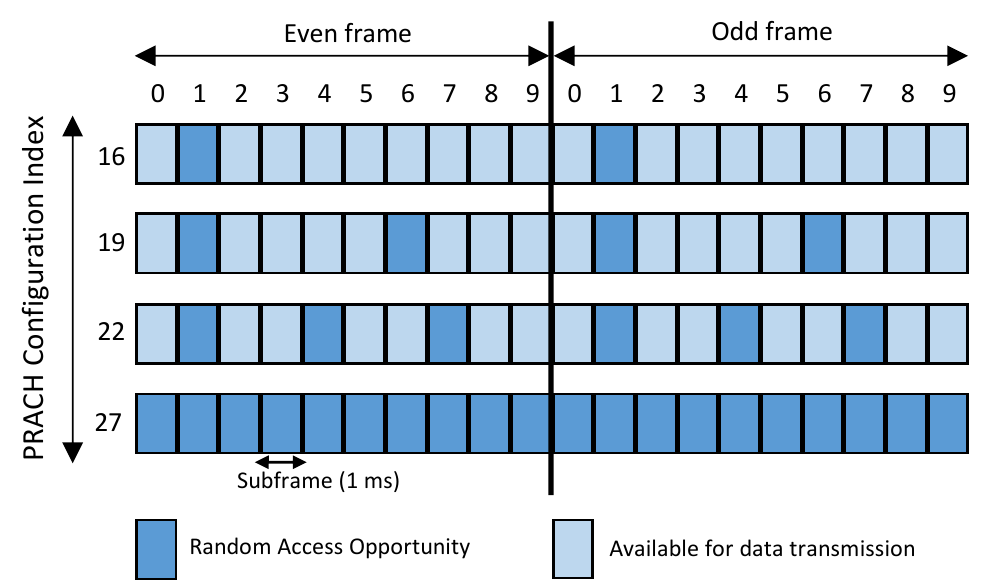}
\caption{Example of different PRACH Configuration Index with $\mu = 0$.}
\label{PRACH-ConfigurationIndex}
\end{figure}

This procedure uses the RACH, which is formed by a periodic sequence of allocated resources in the time-frequency domain, namely RA slots. These slots are reserved in the uplink channel for the transmission of the different access requests. The periodicity of the RA slots is broadcast by the gNB in the SIB, in particular, it is defined by the PRACH Configuration Index parameter \cite{3GPP-38331}. This periodicity ranges from 1 RA slot every 2 frames to a maximum of 1 RA slot per subframe \cite{3GPP-38211}. Fig.~\ref{PRACH-ConfigurationIndex} shows an example of different PRACH Configuration Index values, assuming $\mu = 0$ (same as LTE).

Since RA slots use uplink resources, it is important to maintain a trade-off between the amount of resources dedicated for the RA and the amount of resources available for uplink data transmissions.

\section{Scenario}
\label{Section-Scenario}
In this Section, the simulation setting is described, including the floorplan of the scenario, the activity taking place over this floorplan and the Industry 4.0 applications being used in these activities.

\subsection*{Floorplan}\label{Section-Floorplan}
\begin{figure}[ht]
\centering
\includegraphics[width=\columnwidth]{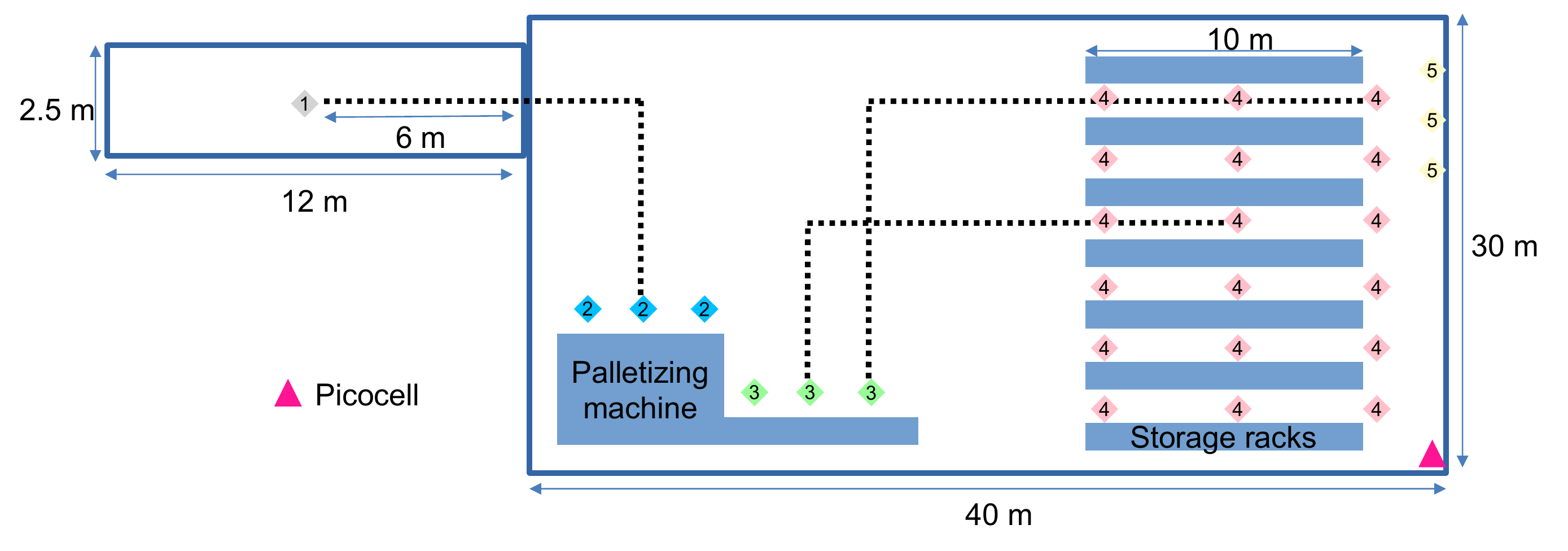}
\caption{Floorplan of the simulated scenario.}
\label{S.floorplan}
\end{figure}

In this paper, a realistic scenario of a small distribution center is reproduced, in order to simulate the behavior of a 5G network supporting different Industry 4.0 applications. The floorplan of the setting is shown in Fig.~\ref{S.floorplan}. The setting is that of a small distribution center with a capacity for a single truck, a shared receiving and shipping dock and a small storage area. The storage area has 7 racks of 10 meters of length, 2 meters of depth and 6 meters of height, separated 2 meters from each other. The distribution center also counts with a single robotic palletizing machine, several AGVs, and is serviced by workers with AR/VR glasses. The floorplan in Fig.~\ref{S.floorplan} also includes a truck trailer that is serviced during the simulation, measuring 12 meters in length, 2.5 meters in width and 3 meters of height (approximately the standard shipping container sizes). The distribution center is also served by one gNB (marked as the magenta triangle). 

\subsection*{Activity}\label{Section-Activity}
The developed simulator can also represent the activity taking place in a distribution center. Fig.~\ref{S.floorplan} shows some numbered marks, each representing one type of location where specific activities take place:
\begin{enumerate}
    \item Truck trailer loading/unloading: AGVs move pallets to/from this point.
    \item Palletizing machine (pallet side): pallets with parcels are either received from or sent to the truck (type 1 point).
    \item Palletizing machine (parcel side): parcels are handled to be palletized and shipped or are processed after being depalletized to be stored in the storage racks.
    \item Storage racks: points where parcels are placed on or retrieved from the storage racks.
    \item Worker access: points where human workers access the installation.
\end{enumerate}

The activity of the different applications is organized around these points. Specifically, four activities are emulated:
\begin{itemize}
    \item Loading/unloading the container: AGV moves between point 1 and a randomly selected point 2. Approximately 12 trips per hour take place with 3 AGVs.
    \item Storing/retrieving parcels: each trip consists of an AGV moving between a randomly selected point 3 and another randomly selected point 4. There are 10 AGVs doing this and each one does 1 round trip per minute.
    \item Human worker activity: workers occasionally retrieve or place parcels on the storage areas, so they perform round trips between a randomly selected point 5 and point 4. The round trip always starts at a point 5. There are 5 workers and each one performs 12 round trips per hour.
    \item Smart tag activity: parcels that are stored in the racks send updates of their location with mMTC. There are 1000 parcels in total.
\end{itemize}

\begin{figure}[ht]
\centering
\includegraphics[width=\columnwidth]{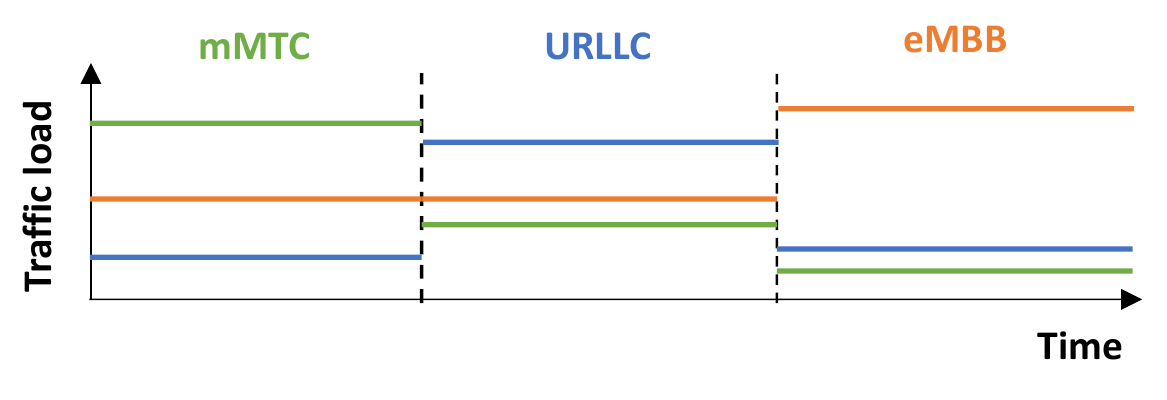}
\caption{Simulated traffic load over time.}
\label{TI.trafficIntensity}
\end{figure}

In this paper, three different time intervals are considered as represented in Fig.~\ref{TI.trafficIntensity}, where in each interval there is one traffic profile that has a high load. More specify, the time intervals correspond to different situations that are performed in the distribution center:

\begin{itemize}
    \item Time interval 1: corresponds to the activity of loading/unloading the container with 3 AGVs, human worker activity with medium traffic load and a massive smart tag activity (1000 simultaneous arrivals).
    
    \item Time interval 2: corresponds to the activity of storing/retrieving parcels with 10 AGVs, human worker activity with medium traffic load and medium smart tag activity (500 simultaneous arrivals).
    
    \item Time interval 3: workers with high traffic load, 3~AGVs storing/retrieving parcels with low traffic load and low smart tag activity (250 simultaneous arrivals).
\end{itemize}

\subsection*{Applications}\label{Section.App}
The applications that run in the simulated distribution center are the following:
\begin{itemize}
    \item AGV messaging: downlink URLLC messages which are generated at a rate of 10 messages per second per AGV, with 64 bytes of data. These messages are critical and will require a low latency and high reliability.
    
    \item AR/VR headsets of workers: each worker has a full-buffer downlink video stream of 8 Mbps for a high traffic activity and 4 Mbps for a medium traffic activity, with 1000 byte packets.
    
    \item Smart Tags: parcels that are processed in the distribution center will be equipped with Smart Tags that regularly send their information to a remote server. These uplink messages occur once per hour, with 64 bytes of data. 
\end{itemize}

\section{Simulator}
\label{Section-Simulator}

To evaluate the performance of the different traffic profiles and their effects when using a static and dynamic NS, two simulators have been used. The first one is used to study the mobile network communications within a distribution center, while the second one is used to evaluate the random access procedure for mMTC devices. Throughout this section, both simulators and their parameters will be described in more details.

\subsection{NS-3 simulator}
\label{Section-NS3}
To study the impact of different slices compared to a static slice, a simulator based on the NS-3 platform has been used, using the 5G-LENA module \cite{5G-LENA}. This module provides a 5G Non-Standalone network and focuses on the new 3GPP NR specifications, which includes features such as numerology support, frequency division multiplexing of numerology, beamforming, among others. In this simulator, a distribution center scenario has been developed with a realistic representation of this particular scenario, where several different logistics activities are done, as explained in Section~\ref{Section-Scenario}.

\subsection*{Network configuration}
The gNB operates with a frequency of 3.7 GHz and a total bandwidth of 20 MHz. 
One transmission/reception omnidirectional antenna is used in both, gNB and UEs, with 15~dBm as downlink transmission power. Regarding numerology, $\mu=0$ has been configured for eMBB and mMTC, whereas $\mu = 2$ has been used for URLLC, which is the highest $\mu$ supported for data channels in FR1, as previously explained in Section~\ref{Section-Numerologies}.

Attending to the slices, two different configurations have been used: 

\begin{itemize}
    \item Static NS: a static slice (baseline) for each type of service with a balance division of network resources.

    \item Dynamic NS: network resources will be appropriately distributed between the different slices according to the traffic load, depending on the activity taking place. This slice will try to maximize the performance of the traffic profile with a peak in each interval and minimize the degradation for the other traffic profiles. 
\end{itemize}

The total bandwidth is divided into three slices, one for each traffic profile. For a static NS, the bandwidth division is configured as 55\% for eMBB, 30\% for URLLC and 15\% for mMTC. On the other hand, when using a dynamic NS, the bandwidth is divided dynamically depending on the time interval:

\begin{itemize}
    \item Time interval 1: 40\% for eMBB, 30\% for URLLC and 30\% for mMTC.

    \item Time interval 2: 30\% for eMBB, 60\% for URLLC and 10\% for mMTC.

    \item Time interval 3: 70\% for eMBB, 20\% for URLLC and 10\% for mMTC.

\end{itemize}

The network parameters alongside traffic parameters on each time interval are summarized in Table \ref{NS3.params}.

\begin{table}[ht]
\centering
\caption{Summary of simulation parameters.}
\begin{tabular}{p{2.83cm}ccc}
\hline
\multicolumn{4}{c}{Network parameters} \\ \hline
Parameter & \multicolumn{3}{c}{Value} \\ \hline
Channel and propagation loss model  & \multicolumn{3}{c}{3GPP 38.901, InF-DH \cite{3GPP-38901}} \\ 
Total bandwidth  & \multicolumn{3}{c}{20 MHz} \\ 
Frequency  & \multicolumn{3}{c}{3.7 GHz} \\ 
Numerology ($\mu$)  & \multicolumn{3}{c}{0 for eMBB and mMTC; 2 for URLLC} \\ 
Transmission direction & \multicolumn{3}{c}{UL for mMTC; DL for eMBB and URLLC} \\ 
Modulation  & \multicolumn{3}{c}{Adaptive} \\ 
Scheduler  & \multicolumn{3}{c}{Round-Robin} \\ 
gNB height  & \multicolumn{3}{c}{10 m} \\ 
gNB transmission power  & \multicolumn{3}{c}{15 dBm} \\ 
UE power control & \multicolumn{3}{c}{3GPP 38.213 \cite{3GPP-38213}} \\ 
MAC to PHY delay  & \multicolumn{3}{c}{2 slots} \\ 
Transport block decode latency  & \multicolumn{3}{c}{100 $\mu$s} \\ 
HARQ feedback delay  & \multicolumn{3}{c}{1 slot} \\ \hline
\multicolumn{4}{c}{Traffic parameters} \\ \hline

Traffic profile & Parameter & Interval & Value \\ \hline
\multirow{4}{*}{eMBB}  & Message size & All & 1000 bytes \\  
                                         & \multirow{2}{*}{Stream rate} & 1, 2 & 4 Mbps \\  
                                         &  & 3 & 8 Mbps \\ 
                                         & Number of nodes & All & 5 \\ \hline

\multirow{4}{*}{URLLC} & Message size & \multirow{2}{*}{All} & 64 bytes \\ 
                                        & Message rate & & 10 Hz \\ 
                                        & \multirow{2}{*}{Number of nodes} & 1, 3 & 3 \\ 
                                        &  & 2 & 10 \\ \hline

\multirow{6}{*}{mMTC}  & Message size & \multirow{3}{*}{All} & 64 bytes \\ 
                                        & Message rate & & 1 per hour \\ 
                                        & Number of nodes & & 1000 \\ 
                                        & \multirow{3}{*}{Simultaneous arrivals} & 1 & 1000 \\ 
                                        & & 2 & 500 \\ 
                                        & & 3 & 250 \\ \hline

\end{tabular}
\label{NS3.params}
\end{table}

\subsection{PRACH simulator for mMTC}
\label{Section-PythonSim}
One of the main drawbacks of the 5G-LENA module \cite{5G-LENA} in NS-3 is that all devices are connected at the beginning of the simulation to the base station in an ideal way. That is, no real RA procedure is performed, the signaling is ideal and it does not consume any radio resources. Furthermore, it does not allow to configure different PRACH Configuration Index values (by default preambles can be sent on any system frame number and subframe number). This is an important aspect that must be taken into account, since the main problem of mMTC comes from the saturation of the RACH as the number of simultaneous devices increases, which results on higher collisions when transmitting the preambles and may cause a device to block if it reaches the maximum allowed RA preamble attempts. 

To overcome this drawback, a new simulator for the RACH has been developed to evaluate the performance of mMTC traffic profile, following the 3GPP standard \cite{3GPP-38211, 3GPP-38321, 3GPP-38331} behavior. This simulator has been implemented on Python and enables to configure different 3GPP parameters for the RACH. In particular, the parameters that can be modified are the following:

\begin{itemize}
    \item PRACH Configuration Index: defines the periodicity of the RA slots. The periodicity ranges between a maximum of one RA slot per subframe to a minimum of one RA slot every two frames. 

    \item Number of available preambles: corresponds to the number of preambles reserved for the contention-based procedure.

    \item \textit{preambleTransMax}: maximum number of preamble attempts for a device before declaring RA failure.

    \item RAR Window Size: time window to monitor RA response.
    
    \item Backoff Indicator: random backoff that is used by the UEs to wait a time when a preamble collision occurs before retrying a new access attempt. This backoff is intended to disperse the access attempts and thus, reduce the probability of preamble collision.

\end{itemize}

\begin{table}[ht]
\centering
\caption{RACH parameters for mMTC.}
\begin{threeparttable}[b]
\begin{tabular}{lccc}
\hline
 Parameter & Interval & Static NS & Dynamic NS \\ \hline
 \multirow{3}{*}{PRACH Configuration Index \tnote{a}} & 1 & & 22 \\
                                        & 2 & 19 & 22 \\
                                        & 3 & & 16 \\ 
 Number of available preambles \tnote{b} & All & \multicolumn{2}{c}{60}\\
 \textit{preambleTransMax} \tnote{c} & All & \multicolumn{2}{c}{10} \\
 RAR Window Size \tnote{c} & All & \multicolumn{2}{c}{5 ms} \\
 Backoff Indicator \tnote{b} & All & \multicolumn{2}{c}{20 ms} \\ \hline
\end{tabular}
\begin{tablenotes}
    \item [a] Refer to 3GPP TS 38.211 \cite{3GPP-38211} for all possible values.
    \item [b] Refer to 3GPP TS 38.321 \cite{3GPP-38321} for all possible values.
    \item [c] Refer to 3GPP TS 38.331 \cite{3GPP-38331} for all possible values.
\end{tablenotes}
\end{threeparttable}
\label{MMTC.params}
\end{table}

Table~\ref{MMTC.params} summarizes the different RACH parameters used in this paper with a static and dynamic slice to evaluate the performance of mMTC. Note that for each time interval (previously defined in Section~\ref{Section-NS3}), the static slice maintains the same parameters, whereas the dynamic slice changes the resources dedicated for the RA procedure (PRACH Configuration Index parameter).

\subsection{Metrics}
In this paper, the following metrics have been considered for eMBB and URLLC:

\begin{itemize}
    \item Throughput (eMBB): measures the quantity of bytes received divided into the simulation time. Note that the simulation time for the throughput calculation is the time elapsed between the first packet transmitted in the transmitter and the latest packet received in the receiver. The ideal situation is when the throughput is equal to the application sent rate, which means that the maximum quality of service is achieved.
    
    \item Reliability (URLLC): it is defined as the number of packets whose latency is below a threshold divided into the total packets transmitted. In this particular case, the latency threshold has been set to 5~ms and it is measured at PDCP layer. The reliability requirement for Mobile Robots must be above $1-10^{-3}$ (99.9\%) \cite{3GPP-22104}. 
    
\end{itemize}

On the other hand, three different metrics have been considered for the evaluation of mMTC with the developed RACH simulator:

\begin{itemize}
    \item Blocking probability: probability that a device reaches the maximum number of transmission attempts (\textit{preambleTransMax}) and is unable to complete an access process.

    \item Average number of preamble retransmissions: measure the average number of preamble retransmissions required to have a success access.

    \item Access delay: time elapsed between the transmission of the first preamble and the reception of the Random Access Response (Msg2) by the mMTC device. Only for devices that do not reach the maximum number of transmission attempts. 
\end{itemize}

\section{Results}
\label{Section-Results}
This section presents the results obtained for the evaluation of each traffic profile and time interval within a distribution center, as defined in previous sections. 

To obtain statistic results, 25 iterations with different seeds have been simulated in the NS-3 simulator, with a duration of 600~seconds in each interval. On the other hand, for the mMTC RACH evaluation, 1000 iterations with different seeds have been performed to obtain the different metrics.

\subsection{URLLC}
\begin{figure}[ht]
\centering
\includegraphics[width=\columnwidth]{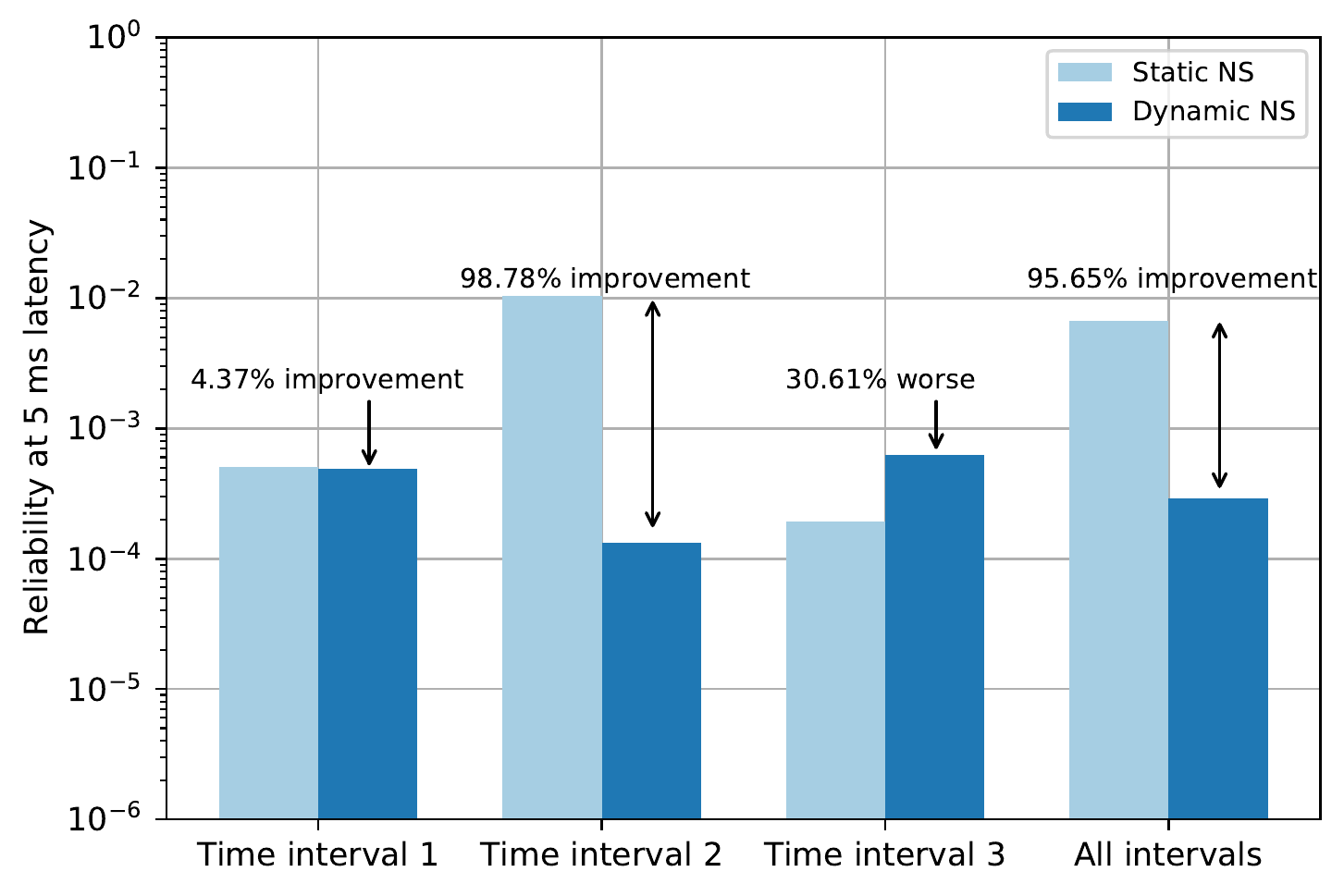}
\caption{URLLC reliability at 5 ms latency with a static and dynamic slice for each time interval.}
\label{Results.Reliability-URLLC}
\end{figure}

Fig.~\ref{Results.Reliability-URLLC} shows the reliability obtained by the AGVs in each time interval and also the combined reliability during all intervals.

Looking at the first time interval, where URLLC traffic is low, it can be noted that the reliability obtained is very similar in both cases (static and dynamic slice). More specify, for a static slice a reliability of $5\cdot10^{-4}$ is obtained; whereas for a dynamic slice the reliability is $4.8\cdot10^{-4}$, with an improvement of 4.37\%. The similarity is due to the use of the same percentage of slice in both cases.

In the time interval 2, where URLLC traffic is very high, it is observed a clearly reduction in the reliability with a static slice compared to a dynamic slice. The static slice obtains a reliability of $10^{-2}$ whereas the dynamic slice obtains an improvement of 98.78\% on the reliability, with a value of $1.3\cdot10^{-4}$. Since a high increment on the traffic of the AGVs is done under this time interval due to the activity taking place, the static slice with a fixed size of 30\% of the total bandwidth (6 MHz) does not provide enough resources for this critical service. Consequently, the latency values are higher (i.e., the probability of receiving packets above the threshold is increased) and therefore, the reliability decreases. On the other hand, since the dynamic slice has incremented its size to 60\% of the total bandwidth (12 MHz), the reliability requirement is guaranteed with a good level (near $10^{-4}$), although the number of AGVs has increased from 3 to 10. 

During the time interval 3, similar to the time interval 1, the intensity of the traffic of the AGVs is low. Under this time interval, the static slice obtains a better reliability ($1.9\cdot10^{-4}$), since the slice size is higher (30\% vs 20\%). On the other hand, the dynamic slice obtains a degradation of 30.61\% on the reliability, with a value of $6.2\cdot10^{-4}$. The fact that the size of the dynamic slice is lower than the static is due to adjust the slice size to maximize the peak traffic profile (eMBB), which will be discussed later. Note that despite using the same amount of resources for the static slice in time intervals 1 and 3, the reliability is better under time interval 3. This is mainly due to the activity taking place and the distance with the gNB (see Fig.~\ref{S.floorplan}). In the time interval 1, the AGVs move between points 1 and 2 (load and unload the container); whereas in time interval 3, the AGVs move between points 3 and 4 (storing/retrieving parcels), which are closer to the base station. Consequently, the propagation losses are higher for those AGVs in the time interval 1.

Taking the values of all time intervals combined, it is clearly noticeable that the dynamic slice performs better (95.65\% improvement) and obtains a reliability of $2.9\cdot10^{-4}$ compared to $6.6\cdot10^{-3}$, which is obtained with the static slice.

\subsection{eMBB}
\begin{figure}[ht]
\centering
\includegraphics[width=\columnwidth]{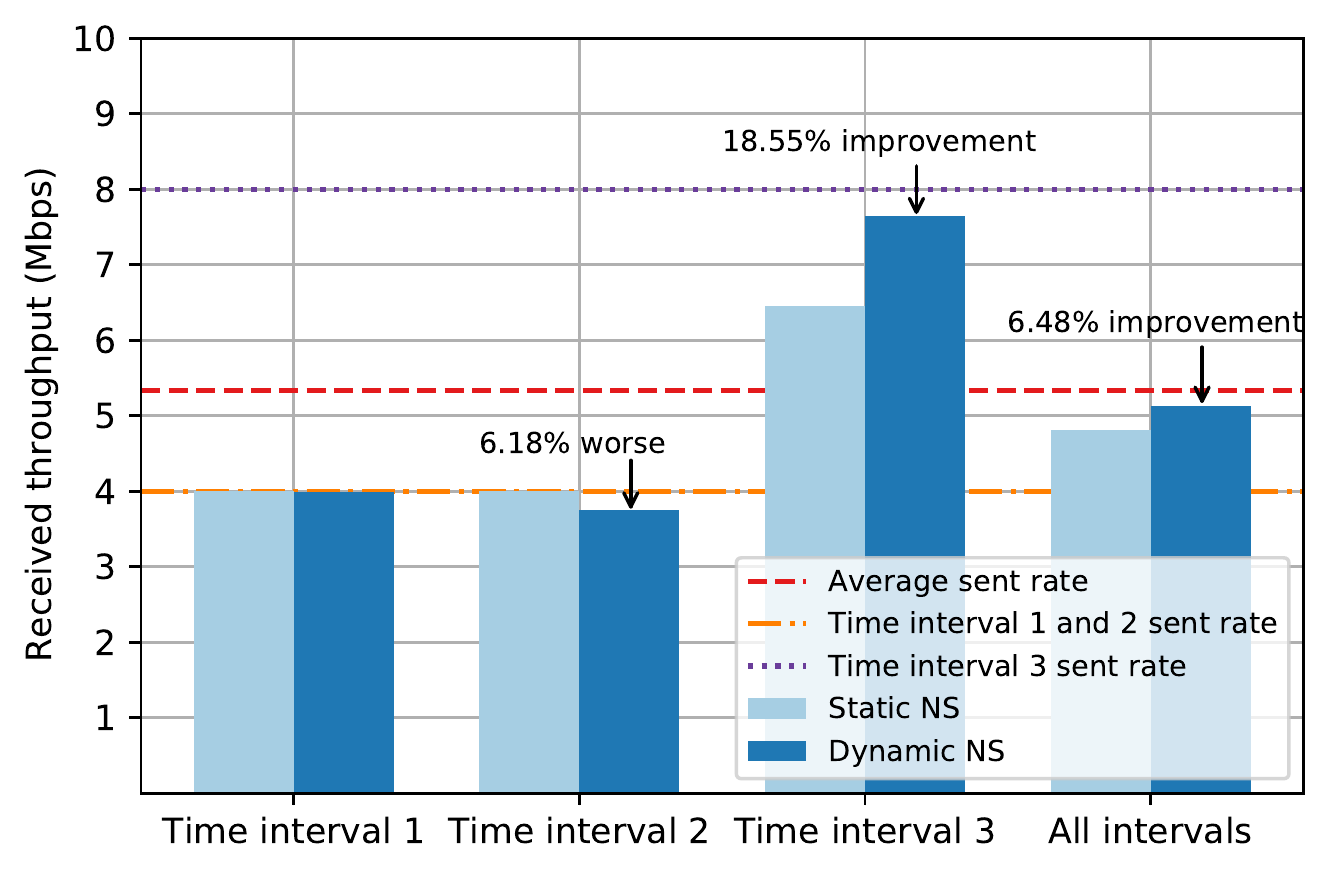}
\caption{Throughput obtained by the workers with a static and dynamic slice for each time interval.}
\label{Results.Throughput-EMBB}
\end{figure}

The average throughput obtained by the workers in each time interval is depicted in Fig.~\ref{Results.Throughput-EMBB}, where the horizontal lines represent the application sent rate in each time interval. 

In the time interval 1, it can be seen that the throughput obtained for both slices (static and dynamic) is equal to the application sent rate, obtaining a value of 4~Mbps. Although the dynamic slice has a reduced bandwidth (40\% of total bandwidth) than the static slice (55\% of total bandwidth), the same throughput is received. This clearly denotes that there is a resource wastage with the static slice; whereas the slice reduction maintains the throughput and this reduction can be used to increase the mMTC slice size, which is the peak traffic in this interval. 

Looking at the second time interval, again, the received throughput is equal to the application sent rate for a static slice. However, since the dynamic slice size has been reduced to 35\% of total bandwidth, a throughput of 3.75~Mbps is obtained. Upon this case, the throughput degradation is minimum (6.18\%) and the resources reduced for eMBB are used to increase the URLLC slice size, which improves the URLLC reliability, as previously seen.

During the time interval 3, the throughput obtained is 6.45~Mbps for a static slice and 7.64~Mbps for a dynamic slice. Under this interval, both slices suffer a throughput degradation, not achieving the application sent rate (8 Mbps). Due to the eMBB traffic peak, more resources are needed, so a static slice cannot fulfill this traffic demand, since its bandwidth is fixed. That is the reason why the throughput is dropped more. On the other hand, although the dynamic slice bandwidth has been increased to 70\% of total bandwidth (14 MHz), the throughput received is slightly reduced but it is very close to the application sent rate, obtaining a throughput improvement of 18.55\% with respect to a static slice.

Taking the values of all time intervals combined, the dynamic slice improves the throughput by 6.48\%, despite the slightly reduction in the time interval 2. In particular, the throughput value obtained is 4.81~Mbps and 5.13~Mbps for a static and dynamic slice, respectively. Note that the throughput of the dynamic slice is very close to the average sent rate (5.33~Mbps).

\subsection{mMTC}
\begin{figure}[ht]
\centering
\includegraphics[width=\columnwidth]{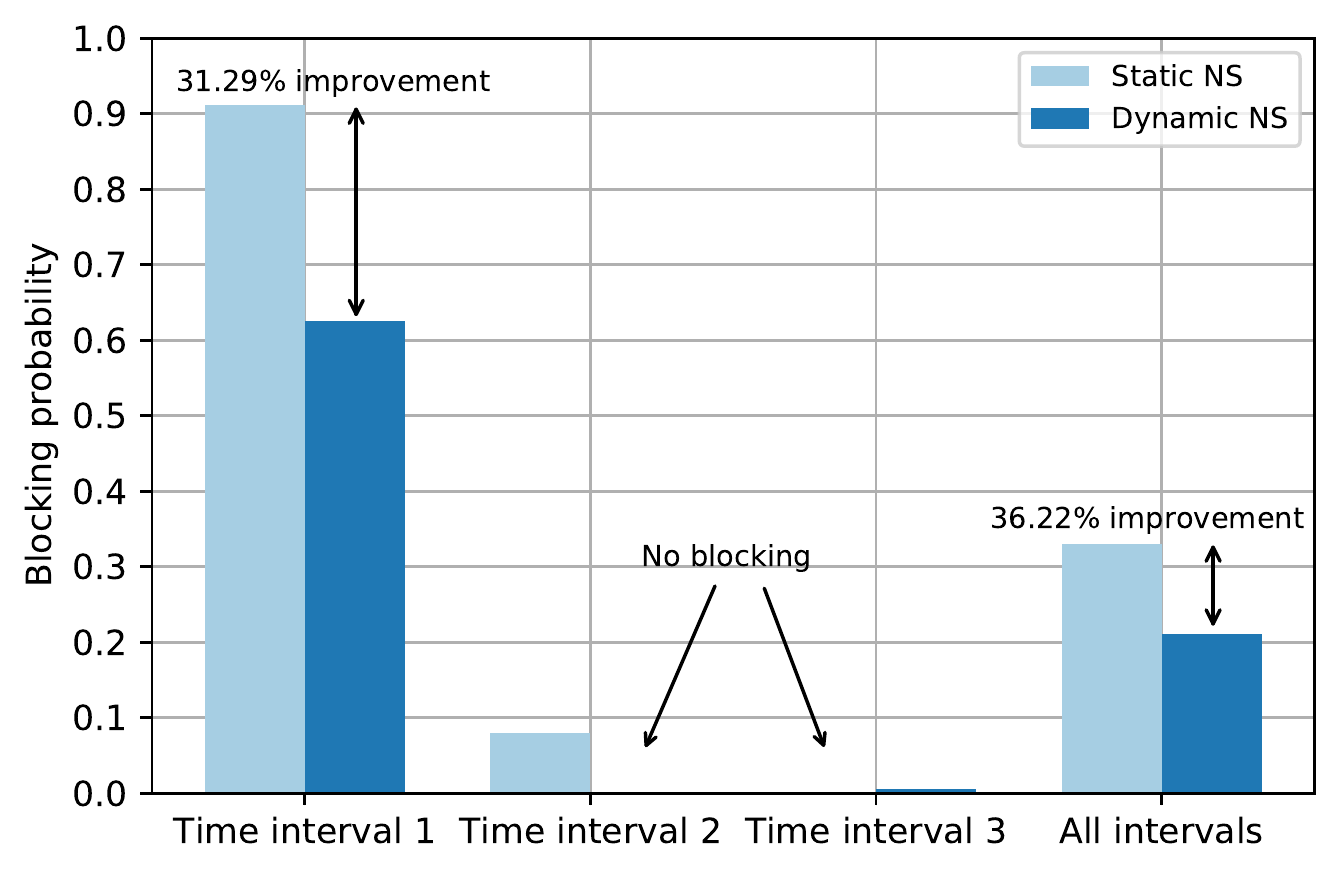}
\caption{mMTC blocking probability with a static and dynamic slice for each time interval.}
\label{Results.Blocking-MMTC}
\end{figure}

The mMTC evaluation has been performed with the developed RACH simulator, as explained in Section~\ref{Section-PythonSim}. Fig.~\ref{Results.Blocking-MMTC} shows the blocking probability obtained by the Smart Tags during the different time intervals. 

During the time interval 1, since there are 1000 simultaneous arrivals, the collisions are very high. A blocking probability of 0.91 is obtained with a static slice, whereas the blocking probability for a dynamic slice is 0.62. Since during this interval the Smart Tags activity is very high, more resources for the RA are dedicated with the dynamic slice (3 RA slots per frame) compared to the static slice (2 RA slots per frame), thus reducing the blocking probability by 31.29\%.

In the time interval 2, since the traffic intensity has decreased (500 simultaneous arrivals), there is no blocking probability with the dynamic slice (it maintains the same RA slots per frame as previous interval), whereas the static slice obtains a blocking probability of 0.08.

The same trend is observed in the time interval 3, where the traffic intensity is low (250 simultaneous arrivals). In this case, there is no blocking probability for the static slice, whereas the dynamic slice (with only one RA slot per frame) obtains a blocking probability of 0.006, which is insignificant. Although the number of RA slots has been decreased for the dynamic slice, the blocking probability is insignificant, but there are more resources available for uplink data transmissions. Therefore, the dynamic slice is more efficient during this interval.

Taking a look in the combination of all intervals, it is clearly visible that the dynamic slice performs better with a reduced blocking probability of 36.22\%, obtaining a probability value of 0.21 compared to 0.33 which is obtained with a static slice.

\begin{figure}[ht]
\centering
\includegraphics[width=\columnwidth]{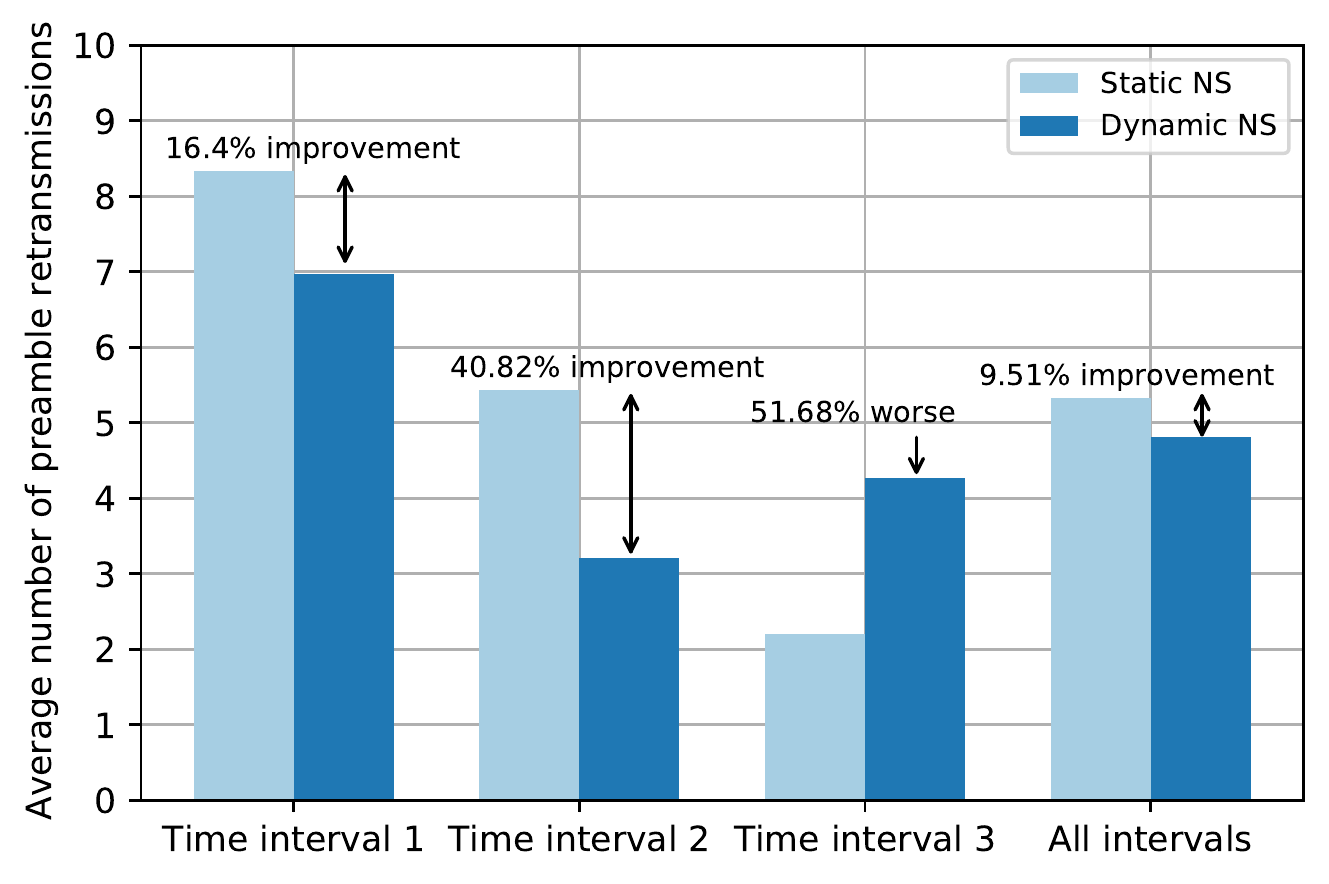}
\caption{mMTC average number of preamble retransmissions with a static and dynamic slice for each time interval.}
\label{Results.AvgPreambleRtx-MMTC}
\end{figure}

The average number of preamble retransmissions needed to have a successful network access is shown in Fig.~\ref{Results.AvgPreambleRtx-MMTC}. As it can be seen, in general, the value is decreased as the time interval is increased, since the simultaneous arrivals are reduced.

In the time interval 1, the average number of preamble retransmissions is close to the maximum allowed value (\textit{preambleTransMax}), which is 10. A value of 8.34 and 6.97 is obtained for a static and dynamic slice, respectively. In this case, the dynamic slice obtains an improvement of 16.4\%.

On the other hand, in the time interval 2, these values are decreased, since there are less simultaneous arrivals and also the blocking probability is very low. The static slice obtains a value of 5.43, whereas the dynamic slice obtains a value of 3.21, which results in a reduction of 40.82\%.

Finally, during the time interval 3, unlike the previous time intervals, it is observed that the dynamic slice obtains a higher value than the static slice, with an increased value of 51.68\%. This is mainly due to the use of less RA slots, which results in more accumulated request on each RA opportunity, and consequently, more collisions occur. Upon this case, the static slice obtains a value of 2.21, wheres the dynamic slice obtains a value of 4.27. 

With the combination of all intervals, although in the interval 3 the static slice performs better, it is compensated on time intervals 1 and 2, so the dynamic slice obtains an improvement of 9.51\% in combination, with a value of 4.82. On the other hand, the static slice obtains a value of 5.32, which is closer to the value obtained with the dynamic slice.

Finally, the average access delay is shown in Fig~\ref{Results.AvgPreambleRtx-MMTC}, where the whiskers represent the upper and lower deviation. Note that the access delay is only measured for the devices that are not blocked. 

In the time interval 1, the static slice obtains an average access delay of 160.92~ms with a lower deviation of 9.07~ms and upper deviation of 11.09 ms. On the other hand, the dynamic slice obtains an average access delay of 124.25~ms with a lower and upper deviation of 6.35~ms and 7.33~ms, respectively. Since more collisions occur with the static slice, achieving a high blocking probability (see Fig~\ref{Results.Blocking-MMTC}), this increments more the delay of the devices that have a successful access. In this case, the dynamic slice reduces the average access delay by 22.78\%.

During the time interval 2, the static slice obtains an average access delay of 100.84~ms with a lower deviation of 9.42~ms and upper deviation of 10.6~ms. On the other hand, the dynamic slice obtains an average access delay of 65.56~ms with a lower and upper deviation of 5.47~ms and 5.07~ms, respectively. Although the blocking probability is very similar with both slices during this time interval (see Fig~\ref{Results.Blocking-MMTC}), the dynamic slice obtains an average access delay reduction of 34.99\%. This reduction is mainly due to the use of more RA slots per frame, which increment the RA opportunities and the devices can achieve a faster network access.

Taking a look in the time interval 3, it can be seen a drastic increment in the delay with a dynamic slice (23.59\%), whereas the static slice reduces the delay. As previously commented, the delay is highly dependent on the periodicity of the RA slots (PRACH Configuration Index value). The higher the periodicity is, the lower the delay is. That is the main reason why the static slice (with 2 RA slots per frame) performs better than the dynamic slice (with 1 RA slot per frame). Upon this interval, the static slice obtains an average access delay of 53.58~ms, with a lower and upper deviation of 5.02~ms and 5.42~ms, respectively. On the other hand, the dynamic slice obtains an average access delay of 94.51~ms, with a lower deviation of 12.21~ms and upper deviation of 13.33~ms.

With the combination of all intervals, it is noted that the access delay is reduced by 9.83\% when using a dynamic slice. Despite the higher delay obtained during the time interval 3, it is compensated with a lower delay when the number of simultaneous arrivals is higher (time intervals 1 and 2).

\begin{figure}[ht]
\centering
\includegraphics[width=\columnwidth]{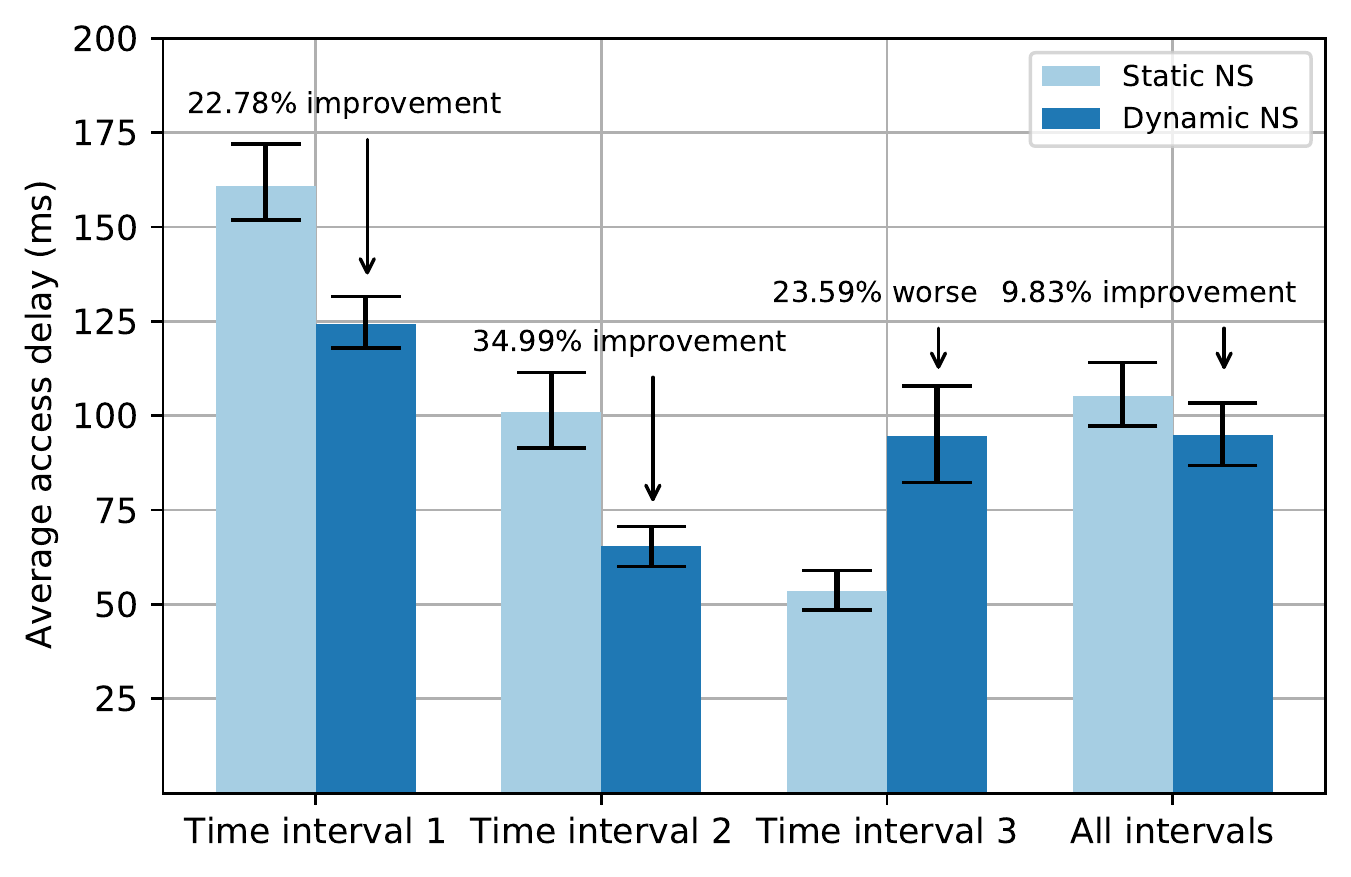}
\caption{mMTC average access delay with a static and dynamic slice for each time interval.}
\label{Results.AccessDelay-MMTC}
\end{figure}

\section{Discussion}
\label{Section-Discussion}

On the one hand, it has been proven that a dynamic slice improves the overall quality of service for the different traffic profiles compared to a static slice. Since the static slice has a fixed size, it guarantees the quality of service only in certain intervals, but not in all. When there is a traffic peak, the static slice does not fulfill the requirements of that traffic, that is, there are not sufficient resources for this eventually traffic intensity. As a consequence, the quality of service decreases and this is especially important for critical services (e.g., AGVs navigation), since a lower quality of service could cause malfunction or accidents within the distribution center. 

On the other hand, a dynamic slice adapts better to the changes on the traffic intensity, maximizing the quality of service of the traffic profile with a peak intensity. When this occur, a slightly degradation of the service is achieved in the other slices, but in combination of all time intervals, the quality of service is improved, since this degradation is negligible compared to the gain obtained when the traffic intensity of this slice has a peak.

\section{Conclusions}
\label{Section-Conclusions}
The objective of this paper has been to study and evaluate two network slicing strategies using the 5G network in a logistics distribution center: the use of a static slice with a balanced division of network resources and the use of a slice that dynamically adjust its size depending on the traffic activity taken place.

The results show that a dynamic slice in fact results in an improved quality of service, especially, under high traffic load. This improvement ranges ranges from 6.48\% to
95.65\%, depending on the specific traffic profile and the evaluated
metric. A static slice performs well when the traffic load is low, since there are sufficient resources to cover the traffic requirements. However, when there is a traffic peak, the static slice does not have enough resources, thus, a reduced quality of service is obtained.

While this study offers a clear view of the potential gains that can be obtained with a dynamic slice in a logistics distribution center, there are some features that remain unexplored yet, such as a scalability analysis and a study of linking the wireless performance with the production performance.
This will be the subject of future work in upcoming experiments with the developed simulator.

\section*{Acknowledgments}
This work has been funded by Ministerio de Asuntos Económicos y Transformación Digital y la Unión Europea - NextGenerationEU within the framework ``Recuperación, Transformación y Resiliencia y el Mecanismo de Recuperación y Resiliencia'' under project MAORI; and by Junta de Andalucía (Consejería de Transformación Económica, Industria, Conocimiento y Universidades, Proyecto de Excelencia) under project PENTA.

\bibliography{Bibliography}
\bibliographystyle{IEEEtran}


\begin{IEEEbiography}[{\includegraphics[width=1in,height=1.25in,clip,keepaspectratio]{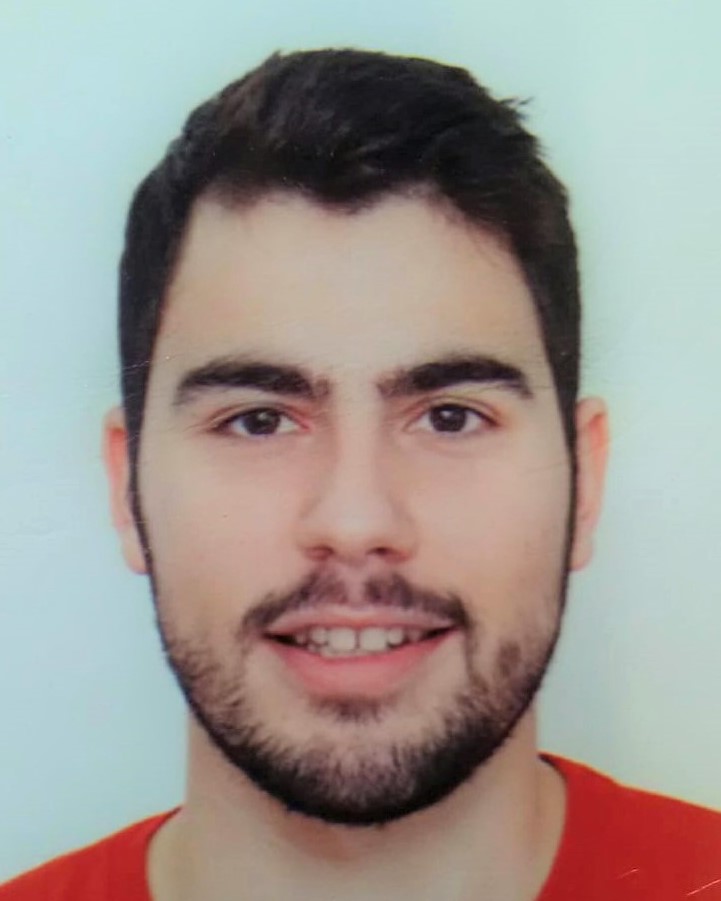}}]{David Segura} received his B.Sc. degree in Telematics Engineering in 2019 and his M.Sc. degree in Telematics and Telecommunication Networks in 2020 from the University of Malaga, Spain. In 2019, he started to work as a research with the Communication Engineering Department, University of Malaga, where he is pursuing a Ph.D. degree in the field of cellular communications.
\end{IEEEbiography}

\begin{IEEEbiography}[{\includegraphics[width=1in,height=1.25in,clip,keepaspectratio]{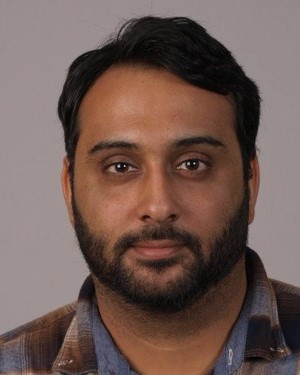}}]{Emil J. Khatib} (Member, IEEE) is a postdoctoral Juan de la Cierva fellow in the University of Málaga. He got a Ph.D in 2017 on the topic of Machine Learning, Big Data analytics and Knowledge Acquisition applied to the troubleshooting in cellular networks. He has participated in several national 
and international 
projects related to Industry 4.0 projects. Currently he is working on the topic of security and localization in industrial scenarios.
\end{IEEEbiography}

\begin{IEEEbiography}[{\includegraphics[width=1in,height=1.25in,clip,keepaspectratio]{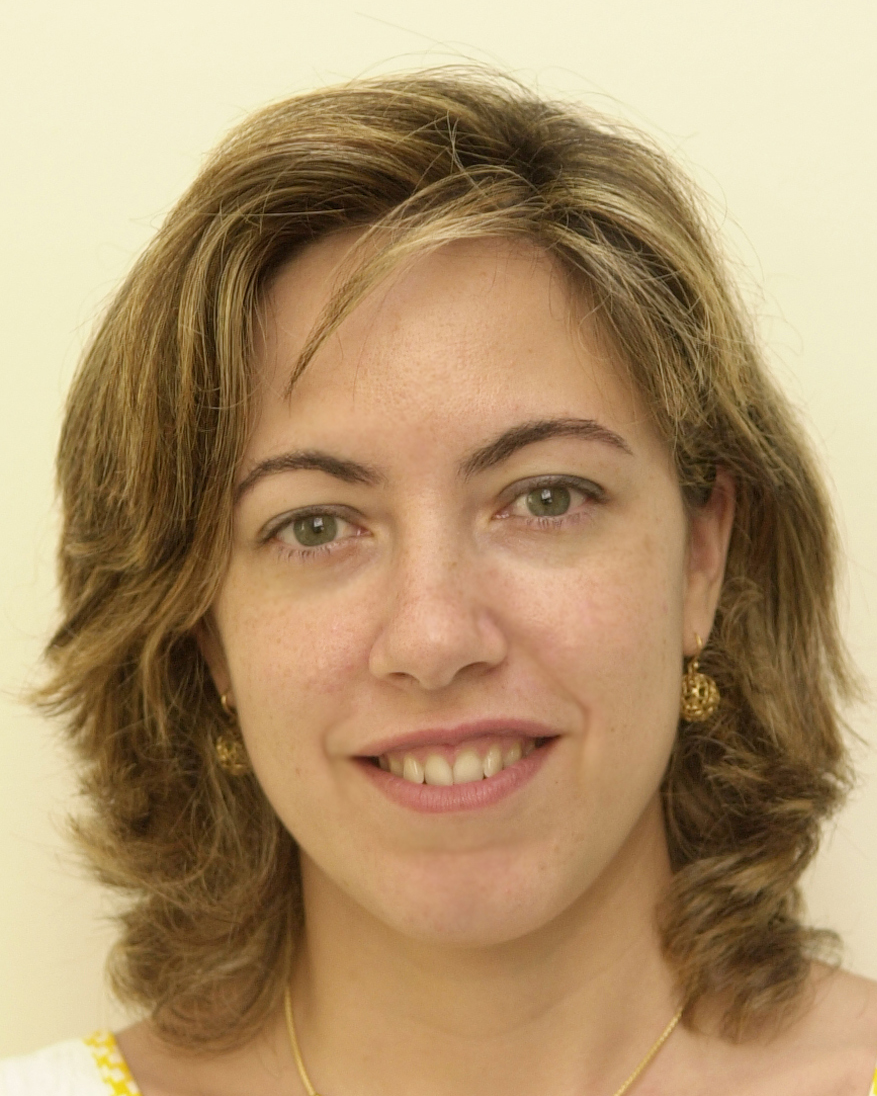}}]{Raquel Barco} is Full Professor in Telecommunication Engineering at the University of Malaga. Before joining the university, she worked at Telefonica (Madrid, Spain) and at the European Space Agency (ESA) (Darmstadt, Germany). As researcher she is specialized in mobile communication networks and smart-cities, having led projects funded by several million euros, published more than 100 papers in high impact journals and conferences, authored 5 patents and received several research awards.
\end{IEEEbiography}

\vfill

\end{document}